\documentclass[singlecolumn,pra,showpacs,preprintnumbers]{revtex4}
\usepackage{graphicx}
\usepackage{amsmath}
\usepackage{graphicx}%

\begin{document}
\title{Comments on \lq\lq Asking Photons Where They Have Been \rq\rq}

\author{Jie-Hui Huang$^{1}$, Li-Yun Hu$^{1}$, Xue-Xiang Xu$^{1}$, Cun-Jin Liu$^{1}$, Qin Guo$^{1}$, Hao-Liang Zhang$^{1}$, and Shi-Yao Zhu$^{2}$}

\affiliation{$^1$College of Physics and Communication Electronics, Jiangxi Normal University, Nanchang 330022, People's Republic of China\\
$^2$Beijing Computational Science Research Center, Beijing 100084,
People's Republic of China}

\begin{abstract}
By using the nonuniform discrete Fourier transform on the center
positions of the symmetric intensity distribution of the output
beam, we recover the vibration information of two mirrors, which is
lost in the analysis of Danan \emph{et al.} in the work [Phys. Rev.
Lett. \textbf{111}, 240402 (2013)]. We believe a photon always
follows continuous trajectories, and a photon has to enter an
interferometer at first and then leave it, if this photon has ever
been inside the interferometer.

\end{abstract}

 \maketitle
Recently, a very interesting experiment was reported in a modified
Mach-Zehnder interferometer system \cite{PRL}. The authors encoded
the photon's path information by slightly
vibrating five mirrors inside the interferometer at different
frequencies, and finally deduce the photons' trajectory through the
power spectrum of the output beam. An interesting phenomenon is found in the second experiment of the work \cite{PRL}, where the power spectrum records the vibration of the mirrors $A$, $B$ and $C$, but without the vibration information of the other two mirrors $E$ and $F$. They thus claimed that \lq\lq the past
of the photons is not represented by continuous trajectories\rq\rq,
and some photons \lq\lq have been inside the nested interferometer,
but they never entered and never left the nested
interferometer\rq\rq \cite{PRL}.

We have to say that such an
interpretation is challenging our intuition too. A basic fact is
that the probabilities of finding photons in the beams $A$ and $B$
in principle sum up to the probability of finding photons in the
beam $E$ (or $F$) all the time. Secondly, our numerical simulation shows that the peaks of $f_A$ and $f_B$ in the second spectrum in Fig. 2 of the reference \cite{PRL} reduce with the increase of the vibration amplitude of the mirror $E$ or $F$ (and disappear finally), accompanied by the appearance of some \lq\lq noisy\rq\rq ~peaks, which is an evidence that the vibration of the mirrors $E$ and $F$ plays a role in the evolution and spatial distribution of the output beam.

We believe the output beam $D$ indeed carries the
vibration information of the two mirrors $E$ and $F$, which can be recovered through the following method. First of all, we use a charge-coupled device (CCD) to replace the quad-cell photodetector to record the
real-time intensity distribution of the output beam. It
is not hard to infer from Eq. (4) in the reference \cite{PRL} that
the intensity has an asymmetric vertical distribution at most time.
However, once the intensity distribution become symmetric in the
vertical direction, we record its center positions and finally Fourier
transform the recorded data.

From the Eq. (4) in the reference \cite{PRL}, we know that
the symmetric intensity distribution occurs when anyone of the
following three conditions is satisfied, (i) $\delta_A=\delta_B$;
(ii) $\delta_C=\delta_B+\delta_E+\delta_F$; (iii)
$\delta_A+\delta_C=2\delta_B+\delta_E+\delta_F$. As a nonlinear function of time
$t$, each one of the three equations has a series of nonuniform
roots. That is to say, the data we sampled in the discrete Fouriour
transform of the center positions of symmetric intensity
distribution (CPSID) are neither uniform nor completely random. We can randomly pick out a part of CPSID
data to make the nonuniform discrete Fourier transform (NUDFT) \cite{NUDFT} and
need not take into acount all records of CPSID.

In our simulation, we randomly choose $N=2.4k$ points among about
$108k$ effective CPSID within $60$ seconds. The Fourier spectrum of
these CPSID data is plotted in Fig. 1 in the range
$[270\text{Hz},340\text{Hz}]$. For simplicity, each involved frequency is
set identical to the one used in the reference \cite{PRL}. Here we see that all
these five frequencies appear in the Fourier spectrum of the CPSID,
accompanied by a few \lq\lq noisy\rq\rq ~peaks. In fact the CPSID
data collected in the current way can be considered as the discrete
points of a continuous function,
\begin{align} \label{eq2}
y(t)=\delta_A-\delta_B+\delta_C+p(\delta_A-\delta_B)(\delta_B-\delta_C+\delta_E+\delta_F)
(\delta_A-2\delta_B+\delta_C-\delta_E-\delta_F) ,
\end{align}
with $p$ a constant. The peaks in the Fourier spectrum of this
continuous funtion are located at the same frequencies as those we
derived from the CPSID data and plotted in Fig. 1(a), but with
different magnitudes and shapes, which helps to explain the
existence of other frequencies besides the five ones we concern, see
Fig. 1(a). Each \lq\lq noisy\rq\rq ~peak can be expressed as a
linear combination of the five frequencies in a simple way, e.g.
$f_1=271\text{Hz}=f_A+f_C-f_E$ and $f_2=310\text{Hz}=f_B-f_E+f_F$.
The low peak of $f_F$ in Fig. 1(a) is related to the coincidence,
$f_F=f_B+f_E-f_A$. By slightly modifying the freqency $f_A$ from
$282\text{Hz}$ to $278\text{Hz}$, we can achieve two strong peaks
for $f_E$ and $f_F$, see Fig. 1(b). Although the strict requirements on the high spatial resolution, fast response and high precision for the
CCD might bring some obstacles for practically
implementing the measurement introduced here, the physics revealed by it theoretically
can not be denied. The photons ever inside the nested interferometer must have entered
and left the nested interferometer. A photon always follows
continuous trajectories.
\begin{figure}
\begin{center}
\includegraphics[scale=0.7]{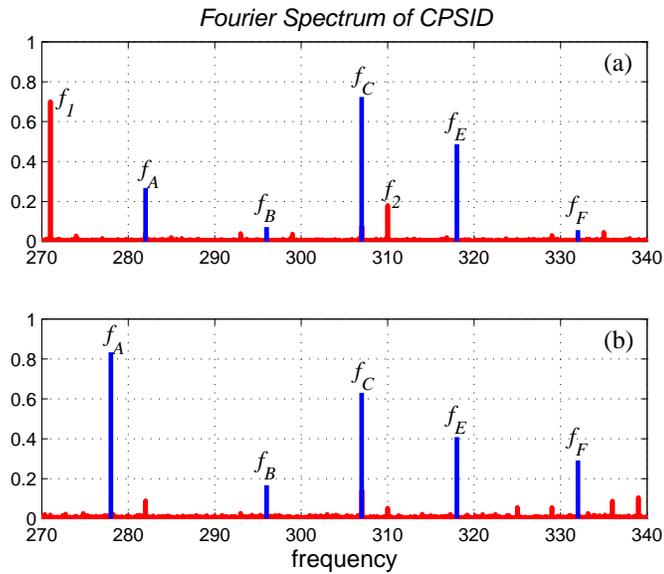}
\end{center}
\caption{(Color online). (a) The vibration frequency of each mirror, $f_i$ $(i=A,B,C,E,F)$, has a peak in the Fourier spectrum of the CPSID. (b) The peak of $f_F$ is enhanced
substantially by slightly changing the frequency $f_A$ from
$282\text{Hz}$ to $278\text{Hz}$.}\label{fig1}
\end{figure}

This work was supported by the national Natural Science Foundation
of China under Grant Nos. 11364022, 11174118 and 11174026.

\end{document}